# 基于区域中心温度场预测的回流焊接优化仿真

隋远， 卜凡洋，邵子龙，闫伟[*]

（山东师范大学信息科学与工程学院，山东 济南 250014）

**摘要**：实现集成电子产品回流焊接前，对回焊炉焊接区域中心温控曲线模拟仿真，有助于选择适当的回焊工艺参数，提高回流焊接工艺整体效率及产品质量。根据热传导规律以及比热容公式，得到焊接区域中心温度曲线关于炉内温度分布函数在传送带位移上的一阶常微分方程，对于温差较小的间隙，使用Sigmoid函数，得到平滑的区间温度过渡曲线；对于温差较大的间隙，利用指数函数和一次函数进行线性组合，追近实际凹函数，从而得到完整的炉内温度分布函数。通过求解常微分方程得到焊接参数，并通过计算预测温度场与真实温度分布数间的均方误差优化模型参数，得到一组符合制程界限的最优工艺参数。同时，根据上述建立的基于区域中心温度场预测方法，针对特定工业生产场景下的实际需求设计了一套回流焊接优化策略：给定温度参数下速度区间预测策略，锡膏融化回流面积最小参数区间预测策略，锡膏融化回流面积左右最对称参数区间预测。仿真结果表明采用该方法得到的温度场预测结果与实际传感器数据高度吻合，具有极强的相关性。该方法可以很好的帮助选择适当的工艺参数，优化生产过程，减少设备调试实践，优化生产产品焊点质量。

**关键词**： 回流焊接；炉温曲线预测；机理模型；常微分方程模型；工艺优化；

**中图分类号**：TH164　　　　　　　　　　　　　　　　　　　　　　**文献标识码**：A

## Optimization simulation of reflow welding based on prediction of regional center temperature field

Sui Yuan，Bu Fan-yang，Shao Zi-long，Yan Wei[(corresponding author*)]

(School of Information Science and Engineering， Shandong Normal University， Jinan Shandong 250014， China)

**ABSTRACT**：Before reflow soldering of integrated electronic products, the numerical simulation of temperature control curve of reflow furnace is crucial for selecting proper parameters and improving the overall efficiency of reflow soldering process and product quality. According to the heat conduction law and the specific heat capacity formula, the first-order ordinary differential equation of the central temperature curve of the welding area with respect to the temperature distribution function in the furnace on the conveyor belt displacement is obtained. For the gap with small temperature difference, the sigmoid function is used to obtain a smooth interval temperature transition curve; For the gap with large temperature difference, the linear combination of exponential function and primary function is used to approach the actual concave function, so as to obtain the complete temperature distribution function in the furnace. The welding parameters are obtained by solving the ordinary differential equation, and a set of optimal process parameters consistent with the process boundary are obtained by calculating the mean square error between the predicted temperature field and the real temperature distribution. At the same time, according to the above established prediction method based on the regional center temperature field, a set of reflow optimization strategies are designed for the actual needs of specific industrial production scenarios: speed interval prediction strategy under given temperature parameters, minimum parameter interval prediction strategy of solder pastes melting reflow area, and the most symmetrical parameter interval prediction of solder paste melting reflow area. The simulation results show that the temperature field prediction results obtained by this method are highly consistent with the actual sensor data, and have strong correlation. This method can help to select appropriate process parameters, optimize the production process, reduce equipment commissioning practice and optimize the solder joint quality of production products.

**KEYWORDS**：reflow soldering; furnace temperature curve; mechanism modeling; ordinary differential equation model; process optimization;

## 1 引言

元件焊接技术是集成电路板等电子产品生产过程中的一项重要工艺，通过加热元器件达到锡膏熔点后，在液态锡表面张力及助焊剂的作用下锡液回流到元器件引脚上形成焊点，进而完成将线路板焊盘和元件焊接成整体的任务。

实际焊接过程中，工业界往往采用回流焊接工艺，即使用回流焊炉设备，将待焊接元器件传送经过多个不同温区，使锡粉完成由固态到液态再到固态的转换，此过程中锡膏在助焊剂等材料的催化下，融化（锡膏熔点：$217^oC$）形成一层薄薄的锡珠,在其表面张力的作用下，聚集在元器件焊点表面，经冷却区制冷凝固，实现焊接。回流焊接作为集成电路板生产中的关键工序，合理的温度曲线设置是保证回流焊接质量的关键[1]。回流焊接的控制实质上是对温度工艺参数的控制[2]，不恰当的温度工艺参数设置会使 PCB 板出现焊接不全、虚焊、元件翘立、焊锡球过多等焊接缺陷，影响产品质量[14]。

目前，关于温度场工艺参数设定大多通过多次重复实验测试进行控制和调整，测试人员使用传感器等设备获取温区温控曲线，凭借操作人员经验观察得出这条曲线反映的能量作用量以及能量作用点，并依次调整温度场工艺参数。产品质量的好坏直接受到操作员经验的影响。现有的仅通过实验获得符合工艺要求的最佳温度的方法不仅低效，而且在每次实验仅只能获得一组炉温曲线的数据，样本量小，测试结果缺乏普适性,极大的浪费了人力财力物力。

为优化回流焊接工艺参数的调控方法，国内外学者进行了相关研究。目前回流焊温度曲线仿真与预测系统主要以 F.Sarvar[6]为代表，提出了部分以机理模型为主要切入点的优化方法[9][10]，徐宗煌[8]提出了一套基于牛顿冷却定律的微分方程的炉温曲线优化模型;席晨曦[11]在微分方程的基础上引入模拟退火算法辅助炉温曲线优化设计；国内的龚雨兵通过数值建模与仿真提出了一种优化的回流焊温度曲线控制；饶庶民[7]在回焊炉温度控制模块的基础上，分析了实际温度场变化规律，使用有限元分析软件对回焊炉内的温度进行仿真分析，可以很好的对温度场进行拟合预测并开发了一套适合回焊炉曲线分析的软件；对于实际回流焊接温度曲线设置，姜海峡[15]论述了回流焊接温度曲线的设置与测试方法，通过对比分析，可调整参数至更加理想的回流焊接温度曲线冯志刚[3]量化了回流焊接工艺参数对温控曲线的影响；张辉华[5]提出了一套面向混装氮气回焊炉的温度曲线控制方案。

为解决传统实验测试方法的弊端，本文采用机理模型进行分析研究，提出一种回流焊接区域中心温度场预测模型，实现区域中心温度场工艺参数的预测优化，并针对特定工业生产场景下的实际需求设计了一套回流焊接优化策略：温度参数已知情况的速度区间预测策略，锡膏融化回流面积最小参数区间预测策略，锡膏融化回流面积左右最对称参数区间预测，能够极大的优化生产过程、节省设备调试时间、优化生产产品焊点质量。

## 2. 基于常微分方程的回流焊接区域中心温度曲线预测模型

### 2.1 模型定义

通常情况下，回流焊炉内部设置若干个小温区，它们从功能上可分成 4 个大温区：预热区、恒温区、回流区、冷却区（如图 1 所示）。某回焊炉内有 11 个小温区及炉前区域和炉后区域（如图 1 所示），每个小温区长度为 30.5 cm，相邻小温区之间有 5 cm 的间隙。其中小温区是指具有加热功能的某一连续加热区间，间隙是指没有加热源的某一连续区间，大温区是指由小温区和间隙组合而成的某一连续区间。实验条件下的回焊炉实际尺寸如表 1 所示。

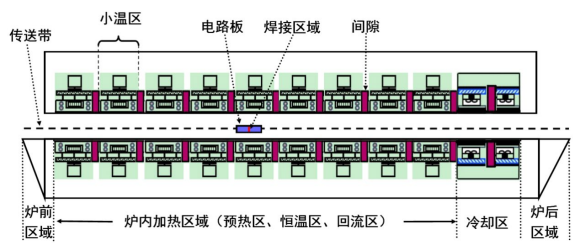

图 1 回焊炉截面示意图

表 1 某回焊炉内含有 11 个小温区及炉前区域和炉后区域的具体尺寸

| 炉内位置 | 区间长度 (cm) | 开始位置 (cm) | 结束位置(cm) |
|---|---|---|---|
| 炉前区域 | 25 | 0 | 25 |
| 小温区 1 | 30.5 | 25 | 55.5 |

| 间隙 1 | 5 | 55.5 | 60.5 |
| 小温区 2 | 30.5 | 60.5 | 91 |
| 间隙 2 | 5 | 91 | 96 |
| 小温区 3 | 30.5 | 96 | 126.5 |
| 间隙 3 | 5 | 126.5 | 131.5 |
| 小温区 4 | 30.5 | 131.5 | 162 |
| 间隙 4 | 5 | 162 | 167 |
| 小温区 5 | 30.5 | 167 | 197.5 |
| 间隙 5 | 5 | 197.5 | 202.5 |
| 小温区 6 | 30.5 | 202.5 | 233 |
| 间隙 6 | 5 | 233 | 238 |
| 小温区 7 | 30.5 | 238 | 268.5 |
| 间隙 7 | 5 | 268.5 | 273.5 |
| 小温区 8 | 30.5 | 273.5 | 304 |
| 间隙 8 | 5 | 304 | 309 |
| 小温区 9 | 30.5 | 309 | 339.5 |
| 间隙 9 | 5 | 339.5 | 344.5 |
| 小温区 10 | 30.5 | 344.5 | 375 |
| 间隙 10 | 5 | 375 | 380 |
| 小温区 11 | 30.5 | 380 | 410.5 |
| 炉后区域 | 25 | 410.5 | 435.5 |

参数可调节范围如表 2 某回焊炉的可调节参数范围所示。

表 2  某回焊炉的可调节参数范围

| | 符号 | 默认参数 | 可调节范围 |
| --- | --- | --- | --- |
| 小温区 1~5 | TT1 | 175ºC | 165ºC~185ºC |
| 小温区 6 | TT2 | 195ºC | 185ºC~205ºC |
| 小温区 7 | TT3 | 235ºC | 225ºC~245ºC |
| 小温区 8~9 | TT4 | 255ºC | 245ºC~265ºC |
| 小温区 10~11 及外界空气温度 | TT5 | 25ºC | 25ºC |
| 传送带速度 | V | 70cm/min | 65 ~100 cm/min |

在设定各温区的温度和传送带的过炉速度后，可以通过温度传感器测试某些位置上焊接区域中心的温度，称之为炉温曲线（即焊接区域中心温度曲线）。实际生产时可以通过调节各温区的设定温度和传送带的过炉速度来控制产品质量。在上述实验设定温度的基础上，各小温区设定温度可以进行±10ºC 范围内的调整。调整时要求小温区 1~5 中的温度保持一致，小温区 8~9 中的温度保持一致，小温区 10~11 中的温度保持 25ºC。传送带的过炉速度调节范围为 65~100 cm/min。

在回焊炉电路板焊接生产中，各温区中心温度场变化应满足一定要求，即制程界限（见表 3）。

表 3  区域中心温度场变化制程界限

| 界限名称 | 最低值 | 最高值 | 单位 |
| --- | --- | --- | --- |
| （1）温度上升斜率 | 0 | 3 | ºC/s |
| （2）温度下降斜率 | -3 | 0 | ºC/s |
| （3）温度上升至 150ºC~190ºC 的时间 | 60 | 120 | s |
| （4）温度大于 217ºC 的时间 | 40 | 90 | s |
| （5）峰值温度 | 240 | 250 | ºC |

2.2 模型建立

假设焊接区域中心看作质点、焊接系数受温度的影响忽略不计、各加热区设定温度即是对应区域炉内温度、焊接过程不考虑热对流。根据热传导规律（1）以及比热容公式（2）对比得出，小温区焊炉内环境温度与焊接中心区域温度的一阶常微分方程(3)：

$$q = -k \frac{dT}{dx} \tag{1}$$

$$q = \frac{Q}{m \cdot dT} \tag{2}$$

$$\frac{dT}{dx} = \frac{-k}{c \cdot v}[T(x) - f(x)] \tag{3}$$

(3)中，$T(x)$ 是炉内环境温度场分布函数，$f(x)$ 是实验测得焊接区域中心温度场分布函数。进一步设 $-\frac{k}{c}$ 为焊接系数 $Q$，得到(4)：

$$\frac{dT}{dx} = \frac{Q}{v}[T(x) - f(x)] \tag{4}$$

2.2.1 回流焊炉炉内温度场分布函数 $T(x)$

由于回焊炉相邻小温区之间炉内环境温度场分布符合 Sigmoid 函数(5)，利用此规律，得到平滑的温度过渡曲线如图 2 所示，曲线两端是恒温区，中间是无加热源的炉内环境温度区，此时 Sigmoid 函数经过平移变换、伸缩变换后得到 $T(x)$ 解析式(6)：

$$S(x) = \frac{1}{1+e^{-x}} \tag{5}$$

$$T(x) = \frac{T_{后温区} - T_{前温区}}{1+e^{-(x-\frac{x_{前}+x_{后}}{2})}} + T_{前温区} \quad (x_{前} \leq x \leq x_{后}) \tag{6}$$

针对小温区与小温区之间温差过大的特殊间隙，利用(4)所述一阶常微分方程推理可知，焊炉内环境温度 $T(x)$ 与焊接中心区域温度

$f(x)$ 温差越小，其焊炉内环境温度 $T(x)$ 的一阶导数越小，因此在 TT4=255℃，TT5=25℃且环境温度也为 25℃的条件下，凹函数存在下降趋势，由于温差的不断缩小，$T(x)$ 的一阶导数也不断减小，此时焊炉内环境温度 $T(x)$ 必然为凹函数。利用位移与焊炉内环境温度 $T(x)$ 的直角坐标系易知坐标 $[x_{前},TT_4]$，$[x_{后},TT_5]$，代入一阶线性函数可得(7)，代入指数函数可得(8)，由(7)(8)可以得到炉内环境温度的线性表达式(9)。

$$\begin{cases} T_1(x) = k(x-x_{后}) + TT_5 \\ k = \dfrac{TT_5 - TT_4}{x_{前} - x_{后}} \end{cases} \quad (7)$$

$$\begin{cases} T_2(x) = A*e^{bx} \\ k = \dfrac{\ln(TT_4) - \ln(TT_5)}{x_{前} - x_{后}} \end{cases} \quad (8)$$

$$A = \dfrac{TT_4}{e^{\frac{\ln(TT_4)-\ln(TT_5)}{x_{前}-x_{后}}x_{前}}}$$

$$T(x) = p \cdot T_1(x) + (1-p) \cdot T_2(x) \quad (9)$$

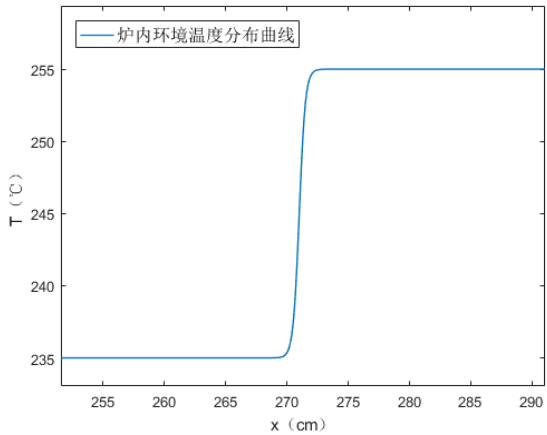

图 2　两端是恒温区，中间是无加热源的炉内环境温度区间过渡曲线

通过遍历参数 $p$，可以确定当 $p = 0.8$ 时方差最小，不同参数 $p$ 下炉温曲线与拟合曲线方差见表 4。

表 4　炉温曲线与拟合数据在不同 $p$ 下方差

| $p$ | 0.6 | 0.7 | 0.8 | 0.9 | 1 |
|---|---|---|---|---|---|
| 方差 | 10161.56 | 6897.62 | 4907.24 | 6333.51 | 6742.85 |

因此，得到炉内环境温度场分布函数如下表 5：

表 5　炉内环境温度场分布函数

| 炉内位置 | 温度场分布函数表达式 |
|---|---|
| 炉前区域 | $TT_5$ |
| 小温区 1~5 | $TT_1$ |
| 间隙 5 | $\dfrac{TT_2 - TT_1}{1+e^{-(x-\frac{197.5+202.5}{2})}} + TT_1$ |
| 小温区 6 | $TT_2$ |
| 间隙 6 | $\dfrac{TT_3 - TT_2}{1+e^{-(x-\frac{233+238}{2})}} + TT_2$ |
| 小温区 7 | $TT_3$ |
| 间隙 7 | $\dfrac{TT_4 - TT_3}{1+e^{-(x-\frac{268.5+273.5}{2})}} + TT_3$ |
| 小温区 8~9 | $TT_4$ |
| 间隙 9~小温区 11 | $0.2 \cdot \dfrac{TT_4}{e^{\frac{\ln(TT_4)-\ln(TT_5)}{339.5-410.5}x_{前}}} \cdot e^{\frac{\ln(TT_4)-\ln(TT_5)}{339.5-410.5}x} + 0.8 \cdot \dfrac{TT_5-TT_4}{339.5-410.5} \cdot (x-410.5) + TT_4$ |
| 炉后区域 | $TT_5$ |

2.2.2　最优焊接系数 $Q$ 预测

利用传感器等工具，测得得焊接区域中心温度曲线 $f(x)$。将 $f(x)$ 与炉内环境温度分布 $T(x)$ 代入公式(4)中，设置各个小温区的温度参数以及传送带过炉速度，并将前一区间预测的 $f(x)$ 的温度末值赋值给后一区间预测的温度初始值，使用四阶龙格库塔法，显式迭代非线性常微分方程，求解出各个位置的解常微分方程的函数值。

通过遍历焊接系数 $Q$，利用作方差等方法挑选出与实验中测得焊接区域中心温度曲线 f(x)最为相近的一组温度曲线,确定并评价最优焊接系数并绘制预测曲线。实验显示得出 $Q = -0.021$ 时方差最小，不同参数 $Q$ 下炉温曲线与拟合曲线方差见表 6。

表 6　炉温曲线与拟合数据在不同 $Q$ 下方差

| $Q$ | 0.0200 | 0.0205 | 0.0210 | 0.0215 | 0.0220 |
|---|---|---|---|---|---|
| 方差 | 3346.01 | 2975.17 | 2912.57 | 3141.50 | 3646.05 |

## 3. 回流焊接方法优化策略

根据 2 提出的基于常微分方程的回流焊接区域中心温度场预测模型，针对特定工业生产场景下的实际需求设计了一套回流焊接优化策略如下：（1）预测设定温度参数下的速度区间；（2）预测锡膏融化时回流面积最小参数区间；（3）预测锡膏融化时回流面积左右最对称参数区间；

### 3.1 预测设定温度参数下的速度区间

各温区温度设定条件下，利用建立的焊接区域中心温度场预测模型，对速度从小到大每隔 0.1cm/min 进行遍历，在所有温区温度确定、焊接中心温度曲线唯一的条件下可以找出符合表 3 制程界限的最大传送带过炉速度，加快工业生产速度。制程界限如下：

对于界限条件（1）、（2），要求升降温速度不超过 3 °C/s，判断是否满足：

$$p = \frac{|T_{n+1} - T_n|}{\Delta t} \leq 3 \quad (10)$$

对于界限条件（3），找到 150℃ 和 190℃ 对应的 $t_1$ 和 $t_2$，判断是否满足：

$$60 \leq t_2 - t_1 \leq 120 \quad (11)$$

对于界限条件（4），找到焊锡熔点温度对应的 $t_1$ 和 $t_2$，判断是否满足：

$$40 \leq |t_2 - t_1| \leq 90 \quad (12)$$

对于界限条件（5），找到 $f(x)$ 的最大值 $T_{\max}$，判断是否满足：

$$240 \leq T_{\max} \leq 250 \quad (13)$$

通过枚举速度，将焊接区域中心温度曲线 $T(x)$ 离散化抽样保存到数组中，判断是否满足制程界限(10)~(13)，记录速度区间并输出。

### 3.2 预测锡膏融化时回流面积最小参数区间

各温区温度设定条件下，利用建立的焊接区域中心温度场预测模型，对速度 $v$、各温区温度设定值 $TT_1,TT_2,TT_3,TT_4$ 进行枚举，计算对应的焊接区域中心温度场变化 $f(x)$，在所有温区温度确定、焊接中心温度曲线唯一、焊接中心温度曲线符合制程界限的条件下，计算温度大于 217℃ 的部分的阴影面积 $s = \int_{x_1}^{x_2} f(x)dx$，经过抽样离散后 $s = \sum_{x_1}^{x_2} f(x)dx$，寻找使得阴影面积最小的各小温区温度以及传送带速度并输出。

### 3.3 预测锡膏融化时回流面积左右最对称参数区间

各温区温度设定条件下，利用建立的焊接区域中心温度场预测模型，对速度 $v$、各温区温度设定值 $TT_1,TT_2,TT_3,TT_4$ 进行枚举，计算对应的焊接区域中心温度场变化 $f(x)$。

通过对焊接中心温度曲线的离散化抽样，利用离散化抽样的方法按时间间隔 t=0.5s 抽样得到焊接中心温度曲线的离散化数组，通过遍历数组的首部与尾部，寻找出 $f(x) = 217℃$ 临界点时刻的 $t_1,t_2$，计算临界点到中心点 $n = \frac{t_1 + t_2}{2}$ 两侧对称区间每对炉温值的方差 $\sum_1^n [f(x_{1n}) - f(x_{2n})]^2$，从中选择即符合制程条件又能使方差最小，对称面积最小的最优解。

## 4. 仿真与分析

为了证明基于区域中心温度场预测的回流焊接模型的有效性，在 Matlab 环境下进行仿真实验测试。炉内环境温度变化 $T(x)$ 和某次实验中测得的焊接区域中心温度场变化 $f(x)$，如图 3 所示，传送带的过炉速度为 70cm/min，炉温各温区温度设定如下：175ºC（小温区 1~5）、195ºC（小温区 6）、235ºC（小温区 7）、255ºC（小温区 8~9）及 25ºC（小温区 10~11）。

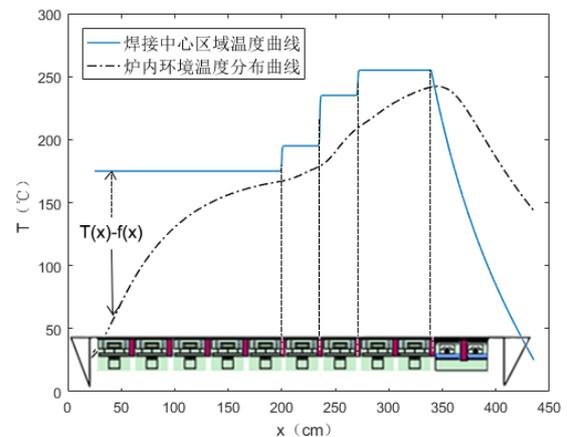

图 3 炉内环境温度变化和某次实验中测得的焊接区域中心温度场变化示意图

图 4 为随机实验中，焊接区域中心温度曲线与不同焊接系数 $Q$ 预测的对比图，发现当 $Q=0.021$ 时，预测效果最佳。当最佳焊接系数 $Q$ 值确定后，任意条件下的焊接区域中心的温度变化情况都可以确定。

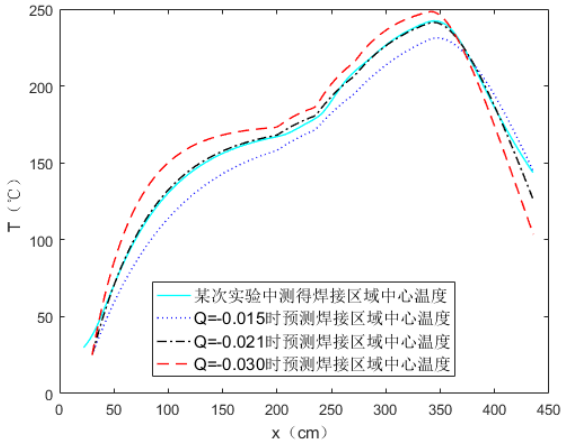

图 4 焊接区域中心温度曲线与不同焊接系数 $Q$ 预测的对比图

为进一步说明本方法的可靠性，使用皮尔逊系数，由表 7 可知当 $Q=0.021$ 时，拟合数据与真实炉温曲线高度吻合，皮尔逊系数高达 99%，表现出极强的相关性，预测结果如图 5 所示。

表 7 实际的炉温曲线与拟合出的炉温曲线在不同 $Q$ 下的方差与皮尔逊相关系数

| $Q$ | 0.0200 | 0.0205 | 0.0210 | 0.0215 | 0.0220 |
|---|---|---|---|---|---|
| 方差 | 3346.01 | 2975.17 | 2912.57 | 3141.50 | 3646.05 |
| Pearson | 0.99923 | 0.99902 | 0.99874 | 0.99797 | 0.99692 |

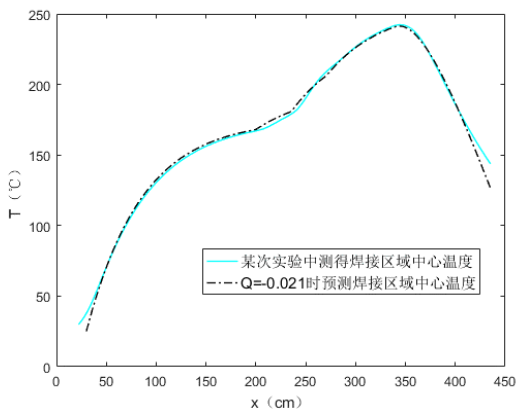

图 5 某次实验焊接区域中心温度曲线与最优焊接系数 $Q=-0.021$ 温度预测的对比图

## 5. 结论

传统回流焊接工艺参数设定时多采用实验测试的方法，存在着低效、耗时、设定结果强依赖经验等问题，因此，本研究采用一种机理模型对该问题进行分析研究，提出一种回流焊接区域中心温度场预测模型，实现焊接区域中心温度场工艺参数的预测优化：根据热传导规律以及比热容公式，得到焊接区域中心温度曲线关于炉内温度分布函数在传送带位移上的一阶常微分方程，对于温差较小的间隙，使用 Sigmoid 函数，得到平滑的区间温度过渡曲线；对于温差较大的间隙，利用指数函数和一次函数进行线性组合，逼近实际凹函数，从而得到完整的炉内温度分布函数。通过求解常微分方程得到焊接参数，并通过计算预测温度场与真实温度分布数间的均方误差优化模型参数，得到一组符合制程界限的最优工艺参数。

同时，根据上述建立的基于区域中心温度场预测方法，针对特定工业生产场景下的实际需求设计了一套回流焊接优化策略：给定温度参数下速度区间预测策略，锡膏融化回流面积最小参数区间预测策略，锡膏融化回流面积左右最对称参数区间预测。

仿真结果表明本方法得到的区域中心温度场预测结果与实际传感器数据高度吻合，具有很强的相关性。因此，本方法可以极大的优化回流焊接生产过程，节省设备调试实践，优化生产产品焊点质量。

## 参考文献：

[1] 陈善. 基于ANSYS仿真的热轧辊瞬态温度场分析[J]. 农业装备与车辆工程，2020, 58(6): 137-140.
[2] 潘开林，周德俭，覃匡宇. SMT再流焊接工艺预测与仿真技术研究现状[J]. 电子工艺技术，2000(5): 185-187.
[3] 冯志刚，郁鼎文，朱云鹤. 回流焊工艺参数对温度曲线的影响[J]. 电子工艺技术，2004，25(6): 243-246，251.
[4] 冯志刚，郁鼎文，朱云鹤.PCB的结构特征对回流温度曲线的影响研究[J].电子元件与材料，2004，(12) :36-39.

[5] 张辉华，黎全英，邴继兵. 混装氮气回流焊接技术研究[J]. 电子工艺技术，2019，40(3): 143-147+156.

[6] Sarvar F，Conway P P.Effective modeling of the reflow pro-cess:use of a modeling tool for product and process design[J].IEEE Transactions on components，packaging and manufacturing technology，1998，(21) :126-133.

[7] 饶庶民. 无铅热风回流炉温度控制系统的设计与研究[D].哈尔滨工业大学，2008.

[8] 徐宗煌,徐剑莆,李世龙,林慧雅.回焊炉电路板焊接炉温曲线优化模型[J].沈阳大学学报(自然科学版)，2021，33(03):279-286.

[9] 孙沛源,吴双琪,吉风池.回流焊炉温曲线优化设计[J].机电信息，2021(14):46-49+52.

[10] 丛铭智,李琪,刘斌,王杰铃,刘靖宇.一种基于机理预测的PCB板回流焊炉温控制方法研究[J].电子技术与软件工程，2020(24):67-69.

[11] 席晨馨. 基于微分方程+模拟退火算法的炉温曲线优化设计[A]. 国家新闻出版广电总局中国新闻文化促进会学术期刊专业委员会.2020年第四届国际科技创新与教育发展学术会议论文集（卷一）[C].国家新闻出版广电总局中国新闻文化促进会学术期刊专业委员会:香港新世纪文化出版社有限公司，2020:3.

[12] 宋会良.基于回流炉温度曲线测试及分析[J].电子测试，2018(13):63-65.

[13] 岳江浩.回焊炉传热温度曲线[J].科学技术创新，2021(12):48-50.

[14] 李恒.回流焊接工艺及常见质量缺陷的改进方法[J].内燃机与配件，2020(08):121-123.

[15] 姜海峡.关于回流焊接温度曲线设置的研究[J].新技术新工艺，2019(08):64-67.